\documentclass[twocolumn,showpacs,superscriptaddress,preprintnumbers,amsmath,amssymb,longbibliography]{revtex4-1}
\usepackage{graphicx}% Include figure files
\usepackage{dcolumn}% Align table columns on decimal point
\usepackage{bm}% bold math
\usepackage{longtable}
\usepackage{hyperref}
\usepackage{lipsum}
\usepackage{braket}
\usepackage{physics}
\usepackage{setspace}
\usepackage{color}
\usepackage{fancyhdr}
\usepackage{titlesec}
\usepackage{float}
\usepackage{upgreek}
\usepackage{array}
\usepackage{epstopdf}

\usepackage[utf8]{inputenc}

\usepackage{wrapfig}
\usepackage{amsmath,amsthm,amssymb}
\usepackage[table,xcdraw]{xcolor}

\titleformat*{\section}{\LARGE\bfseries}
\titleformat*{\subsection}{\Large\bfseries}
\titleformat*{\subsubsection}{\large\bfseries}

\begin{document}
\preprint{APS/123-QED}
\title{Reliable magnetic domain wall propagation in cross structures for advanced multi-turn sensor devices}

\author{B. Borie}%
\affiliation{Institut f\"ur Physik, Johannes Gutenberg-Universit\"at Mainz, Staudinger Weg 7, 55128 Mainz, Germany.}%
\affiliation{Sensitec GmbH, Hechtsheimer Straße 2, Mainz D-55131, Germany.}%

\author{M. Voto}%
\affiliation{Departamento de F\'isica Aplicada, Universidad de Salamanca, Plaza de la Merced s/n, E-37001, Salamanca, Spain.}%

\author{L. Lopez-Diaz}%
\affiliation{Departamento de F\'isica Aplicada, Universidad de Salamanca, Plaza de la Merced s/n, E-37001, Salamanca, Spain.}%

\author{H. Grimm}%
\affiliation{Sensitec GmbH, Hechtsheimer Straße 2, Mainz D-55131, Germany.}%

\author{M. Diegel}%
\affiliation{Leibniz-Institut für Photonische Technologien, Albert-Einstein-Straße 9, 07745 Jena, Germany.}%
\email{Email: klaeui@uni-mainz.de, roland.mattheis@leibniz-ipht.de}

\author{M. Kl\"aui}%
\affiliation{Institut f\"ur Physik, Johannes Gutenberg-Universit\"at Mainz, Staudinger Weg 7, 55128 Mainz, Germany.}%

\author{R. Mattheis}%
\affiliation{Leibniz-Institut für Photonische Technologien, Albert-Einstein-Straße 9, 07745 Jena, Germany.}%

\date{\today}% It is always \today, today,

\begin{abstract}

We develop and analyze an advanced concept for domain wall based sensing of rotations. Moving domain walls in n closed loops with n-1 intersecting convolutions by rotating fields, we can sense n rotations. By combining loops with coprime numbers of rotations, we create a sensor system allowing for the total counting of millions of turns of a rotating applied magnetic field. We analyze the operation of the sensor and identify the intersecting cross structures as the critical component for reliable operation. In particular depending on the orientation of the applied field angle with the magnetization in the branches of the cross, a domain wall is found to propagate in an unwanted direction yielding failures and counting errors in the device. To overcome this limiting factor, we introduce a specially designed syphon structure to achieve the controlled pinning of the domain wall before the cross and depinning and propagation only for a selected range of applied field angles. By adjusting the syphon and the cross geometry, we find that the optimized combination of both structures prevents failures in the full sensor structure yielding robust operation. Our modeling results show that the optimized element geometry allows for the realization of the sensor with cross-shaped intersections and operation that is tolerant to inaccuracies of the fabrication.

\end{abstract}

\maketitle

\section{Introduction}

Magnetic domain walls in soft ferromagnetic conduits have been under research since the mid-1960s \cite{Spa66, Spa70, McM97, MK08, All02, All05, Kle08, Oma14, Par08, Don10, Die04, Die07, Die09, Mat12}. Proposals using domain walls (DWs) in soft magnetic nanostructures \cite{McM97, MK08} as key elements have been put forward to realize magnetic logic gates \cite{ All02, All05, Kle08, Oma14}, memory devices such as race track memory \cite{Par08}, magnetic bead transportation \cite{Don10}, or turn sensors with storage capability (multi-turn counter) \cite{Die04, Die07, Die09, Mat12}. 
The first industrial device realization is a multi-turn counter, which can sense and store the number of turns with a true-power-on functionality. This sensor can count from 0 to 16 and back, it is manufactured by Novotechnik and already commercially available [RSM-2800 Multiturn and RMB-3600 Multiturn, \cite{Novo}]. 
The concept of this kind of sensors is the generation and movement of 180$^\circ$ DWs along Giant Magnetoresistance nanowires in an open-loop-spiral-like geometry (see Fig. \ref{fig1}-a). Many industrial applications require a significant number of counts (from hundreds up to millions). However, the scaling of this particular geometry remains limited, and these number of counts can thus not be achieved.

A different approach was recently proposed based on using closed loops with a different number of cusps directed toward the center of the loop \cite{Mat12}. However, an issue inherent to the cusp geometry is the double width caused by the merging of two nanowires at this position. This characteristic imposes a reduction of the maximum sensed field before unwanted and uncontrolled random nucleation is initiated in this wider part. The cusp geometry thus narrows the field operating window (FOW) of the sensor, which is defined by the difference between the maximum magnetic field above which unwanted DW nucleation occurs and the minimum field necessary for a reliable transport of the DW through out the whole structure. 
Compared to a perfect straight stripe, the minimum field necessary for propagation through a cusp is also significantly increased. Thus the FOW is narrowed by an increased minimum and reduced maximum operating field, making this device geometry challenging for real use.

Due to the limitations imposed by the reduced FOW, there is a need for the development of alternative concepts. A possible different concept uses the combination of intersecting closed loops capable of counting coprime numbers of turns.  This concept includes a different geometrical feature, namely a cross of nanowires to allow the intertwining of loops (Fig. \ref{fig1}-b)\cite{Mat14}. This geometry fundamentally allows for a much improved multi-turn counter. We name such a structure a n-CL, where n is the number of closed loops. For example, Fig. \ref{fig1}-b shows a 3-CL device with 3 closed loops and 2 crossings that can count 4 turns. Every n-CL contains a minimum of 2 and a maximum of 2n-2 of DW for sensing.

An intrinsic characteristic of the closed loop structure (Fig. \ref{fig1}-b), and arguably the essential one, is the possibility to automatically reset back to counting from 0 after achieving the maximum number of turns offered by the architecture. 
Using this resetting mechanism with the concept of coprime number counting permits to achieve counting of a much larger number of turns.
This type of counting is allowed by positioning several n-CLs next to each other with n = 3, 5, 7, and so forth with n being the coprime number. The results of individual structures are combined to enable counting to large numbers.
The simultaneous use of a 3-CL, a 5-CL and a 7-CL (Table \ref{fig0}) yields different sequences of output states allowing a maximum count of 3 x 5 x 7 = 105 turns, which is already more than any open-loop structure (Fig. \ref{fig1}-a) could ever do. The method is scalable and with 7 different n-CLs (n = 5, 7, 9, 13, 16, 17, and 19), the number of turns available is  already 5 x 7 x 9 x 13 x 16 x 17 x 19 = 21 Mio. turns.

\begingroup
\squeezetable
\begin{table}[H]
\centering
\begin{tabular}{|c|c|c|c|}
\hline
\rowcolor[HTML]{68CBD0} 
\textbf{Maximum number of turns}          & \textbf{3-CL} & \textbf{5-CL} & \textbf{7-CL} \\ \hline
\rowcolor[HTML]{FFFFFF} 
\textbf{State of the device after 1 turn} & \textbf{1}    & \textbf{1}    & \textbf{1}    \\ \hline
\rowcolor[HTML]{FFFFFF} 
\textbf{2 turns}                          & \textbf{2}    & \textbf{2}    & \textbf{2}    \\ \hline
\rowcolor[HTML]{FFFFFF} 
\textbf{3 turns}                          & \textbf{3}    & \textbf{3}    & \textbf{3}    \\ \hline
\rowcolor[HTML]{FFFFFF} 
\textbf{4 turns}                          & \textbf{1}    & \textbf{4}    & \textbf{4}    \\ \hline
\rowcolor[HTML]{FFFFFF} 
\textbf{5 turns}                          & \textbf{2}    & \textbf{5}    & \textbf{5}    \\ \hline
\rowcolor[HTML]{FFFFFF} 
\textbf{6 turns}                          & \textbf{3}    & \textbf{1}    & \textbf{6}    \\ \hline
\rowcolor[HTML]{FFFFFF} 
\textbf{16 turns}                         & \textbf{1}    & \textbf{1}    & \textbf{2}    \\ \hline
\rowcolor[HTML]{FFFFFF} 
\textbf{31 turns}                         & \textbf{1}    & \textbf{1}    & \textbf{3}    \\ \hline
\rowcolor[HTML]{FFFFFF} 
\textbf{91 turns}                         & \textbf{1}    & \textbf{1}    & \textbf{7}    \\ \hline
\rowcolor[HTML]{FFFFFF} 
\textbf{106 turns}                        & \textbf{1}    & \textbf{1}    & \textbf{1}    \\ \hline
\end{tabular}

\caption{Table summarizing the output of the devices for different count numbers.}
\label{fig0}
\end{table}

\endgroup

This closed loop multi-turn sensor thus opens up additional fields of applications where the open multi-turn counter would be inefficient, such as, for example, highly sensitive angle detection via pole wheels. 

However, a fundamental problem for this sensor operation arises at the eventuality that a DW does not propagate straight through the cross but takes a turn and thereby changes its path leading to a counting failure event. This event is particularly likely if the external field direction that drives the DW through the cross is not well controlled. Therefore, more sophisticated device geometries than the one shown in Fig. \ref{fig1}-b are needed for reliable device operation, robustness and fault tolerance.

In this paper, we present a novel sensor architecture that reliably counts large numbers of complete turns of a rotating applied magnetic field. The structure is designed to comprise a new syphon shaped element in addition to the cross-shaped intersections of nanowires \cite{Die15}. This allows for the desired control of the propagation direction of the DW under the application of a rotating field. First, we introduce the different possible states that can be present in the cross structure and explain the necessity for a syphon structure.
Second, the angular dependence of the crossing and the syphon are separately simulated to extract the critical points of the two geometries. Finally, the results are combined to observe the behavior of a complete structure. We identify the three key parameters that allow us to gauge the reliability of the structure, and optimized devices are demonstrated for further improvements of the concept.

\section{Concept for a DW based multi-turn counter with crosses and syphons elements}
                                                                                                                                                                                                                                                                                                                                 
\begin{figure}[H]
\centering
\includegraphics[scale=0.4]{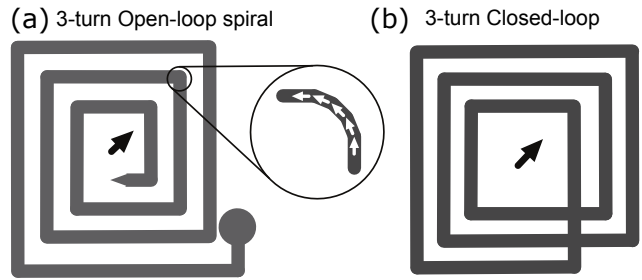} 
\caption{ (a)  A 3-turn-open-loop spiral with DW nucleation pad and a pointed end. The enlargement shows the polygon-shaped corner with 22.5$^\circ$ kink angles usually used \cite{Novo}.  (b) A 3-turn closed-loop structure with two cross-shaped intersections. Note that, in a real device, the distance between two crosses is expected to be much larger than the wire width effectively decoupling the crosses.}
\label{fig1}
\end{figure}

To overcome the limitations caused by the cusp geometry, we propose the new concept, schematically shown in Fig. \ref{fig1}-b. Here the inner and outer end of the spiral of Fig. \ref{fig1}-a are connected via a nanowire generating a cross-shaped intersection. 
In magnetically soft wires, the nucleation field for a DW follows a simple Stoner-Wohlfarth like model \cite{Bor17} and it is thus an inverse function of the stripe width. A cross leads to only a $\sqrt{2}$-fold increase of the width of the nanowire in its diagonal as compared to a 2-fold increase with the cusp geometry \cite{Mat12}. Quantitatively, with a 30 nm thick and 300 nm wide stripe of Ni$_{81}$Fe$_{19}$, the nucleation field is therefore improved by at least 40$\%$ by utilizing a cross structure.

The propagation of DWs in straight and curved wires is well established \cite{MK08}, and the motion of a DW through the cross has been studied \cite{Lew09, Lew10}. However, the details of the reliability of this process are still under active investigation due to the complex dynamics involved in the process. In this section, dedicated to the explanation of the concept of the sensor mechanism, we used a transverse domain wall type as a simple representation of a DW.

\begin{figure}[H]
\centering
\includegraphics[scale=0.4]{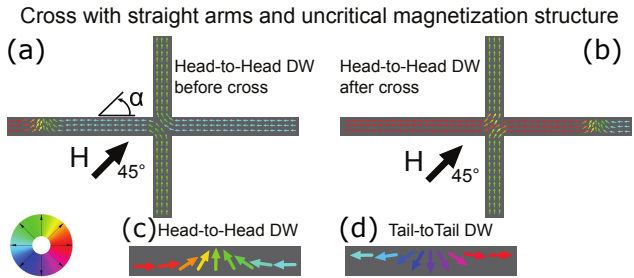} 
\caption{Sketch of a DW configuration before and after moving across a cross with an energetically favored magnetization state of the vertically oriented arm. $\alpha$ is the angle between the applied field direction and the horizontal direction. (a) DW positioned before the cross in the left arm. A field H at $\alpha$ = 45$^\circ$ moves the DW towards the cross. (b) DW positioned in the right arm after moving through the cross.  (c) Schematic representation of a Head-to-Head DW, and (d) a Tail-to-Tail DW. The color code indicates the in-plane magnetization direction as shown by the color wheel.}
\label{fig2}
\end{figure}

The quasi-static states before and after traveling through the cross are represented in Fig. \ref{fig2}. They illustrate the process of a moving DW in the horizontal arm from the left to the right side under conditions where the magnetization of the vertical arm is in the energetically favored state, i.e. parallel to some component of the applied field (see Fig. \ref{fig2}-a). 
Since both branches have the same width, the magnetization structure in the cross is stray-field free. When the DW is to the left of the cross (Fig. \ref{fig2}-a), there is a continuous magnetic flux from bottom to left and from right to up. The magnetization is along a 135$^\circ$ direction in the core of the cross. When the DW has moved through the cross, the core magnetization is rotated by 90$^\circ$ in the clockwise direction forming a stray-field-free state (Fig. \ref{fig2}-b) again.

\begin{figure}[H]
\centering
\includegraphics[scale=0.4]{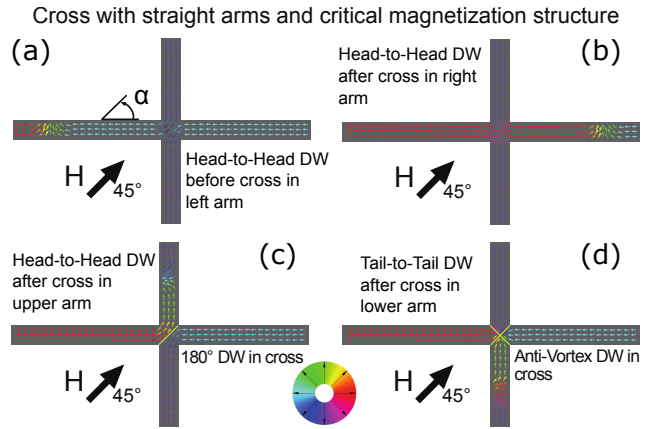} 
\caption{ Sketch of the DW configuration before and after moving through a cross with the possible resulting magnetization structures. $\alpha$ is the angle between the applied field direction and the horizontal direction. The applied field can remagnetize the horizontal arm or the vertical arm. (a) The DW is positioned before the cross in the left arm. A field H inclined at 45$^\circ$ moves the DW towards the cross. For $\alpha$ = 45$^\circ$ the DW can be positioned in any of the three arms after moving through the cross: (b) DW placed in the right arm. (c) DW placed in the upper (wrong) arm and (d) DW positioned in the lower (wrong) arm.  In (a) and (b), the cross is DW-free. In (c), a 180$^\circ$ DW and in (d) an Anti-Vortex DW is located in the center of the cross. Additionally, in (d) the Head-to-Head DW must transform into a Tail-to-Tail DW during the movement through the cross. The in-plane magnetization direction is indicated by the color code visible in the color wheel.}
\label{fig3}
\end{figure}

In Fig. \ref{fig3}, we present the second possible starting configuration. In this case, the magnetization of the vertical arm is in a high energetic state as the magnetization is anti-parallel to some component of the applied field.
Nevertheless, the magnetization pattern of the cross is stray field free, and the core magnetization is directed in a 225$^\circ$ direction as shown in Fig. \ref{fig3}-a. 
As a result of this configuration, the magnetization in the top, bottom, and right arm are in a higher energetic state. Therefore the DW could travel horizontally or vertically (up or down) resulting in the three states shown in Fig. \ref{fig3}-b, c, and d. 
Only the movement in the horizontal path (Fig. \ref{fig3}-b) yields a stray field free configuration (switching of the core magnetization by 90$^\circ$ from  225$^\circ$ to  315$^\circ$). Any propagation of the DW into the up or down arms results in a magnetization structure with a 180$^\circ$ DW (Fig. \ref{fig3}-c) or an Anti-Vortex state (Fig. \ref{fig3}-d) in its center. 

From an energetic point of view, the energy barrier to overcome in the final states in Fig. \ref{fig3}-c and d is higher than in the sole horizontal depinning in Fig. \ref{fig3}-b. This is due to the required energy to create an additional DW at the cross. From this simple picture, we intuitively expect a preferred DW motion in the horizontal direction toward the stray field-free state in Fig. \ref{fig3}-b instead of the more energetic states in Fig. \ref{fig3}-c and d.
Despite those different energy barriers for the different final states, if the applied field exhibits a large component along the vertical direction, the DW is likely to move vertically resulting in a failure of the operation of the sensor device. Therefore, it is desirable to constrain the DW propagation to angles around $\alpha$ = 0$^\circ$ respectively 180$^\circ$.

\begin{figure}[H]
\centering
\includegraphics[scale=0.4]{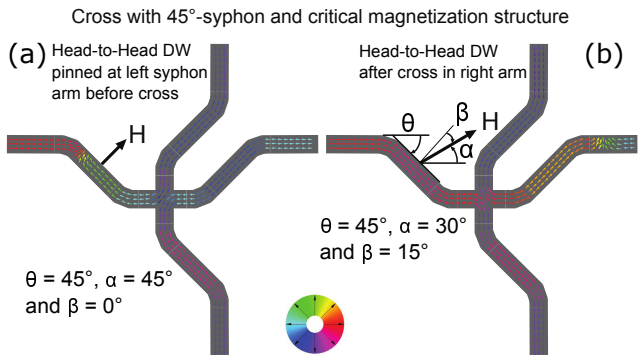} 
\caption{ Sketch of the DW motion through a cross combined with a 45$^\circ$ syphon. $\theta$ is the angle of the syphon arm with respect to the horizontal direction, $\alpha$ is the field angle with respect to the horizontal direction, and $\beta$ is the angle between the perpendicular direction with the wire containing the DW and the applied field direction. (a) A DW is pinned at the beginning of the syphon in the left arm for H inclined at 45$^\circ$, resulting in no longitudinal component to move the DW into the syphon ($\beta$ =  0$^\circ$). (b) The DW moved through the cross with H inclined at 30$^\circ$ ($\beta$  = 15$^\circ$).}
\label{fig4}
\end{figure}

However, in a simple closed-loop multi-turn counter, the DWs are moved by a rotating magnetic field. We consider in this case a clockwise rotation of the field. The motion in the horizontal nanowire starts depending on the natural pinning strength for a Head-to-Head DW in the wires at field directions around $\alpha$ = 70$^\circ$ – 80$^\circ$ because the horizontal component of the applied field is sufficient to depin. Under these conditions, the DW would not reliably travel from the left arm to the right arm as desired, but instead into one of the vertical arms as depicted in Fig. \ref{fig3}-c and d, resulting in a counting error of the sensor.  To prevent this failure, we introduce a novel syphon-like geometry (Fig. \ref{fig4}).  

This syphon has the aim to pin the DW in its arms until the field angle is favorable for a propagation only in the horizontal direction. For example, in Fig. \ref{fig4}-a, with $\alpha$ = 45$^\circ$ and $\theta$ = 45$^\circ$ (syphon arm angle), $\beta$ the angle with the perpendicular to the syphon arm is 0$^\circ$ thus the applied field is perpendicular to the syphon arm, and the DW is pinned in the arm. In Fig. \ref{fig4}-b, with the applied field angle $\alpha$ = 30$^\circ$, the angle $\beta$ = 15$^\circ$ thus there exist a component of the field along the syphon arm and the DW can depin. At field angles lower than $\alpha$ = 45$^\circ$, a larger torque is provided in the horizontal direction than in the vertical thus favoring horizontal propagation through the cross as depicted in Fig. \ref{fig4}-b.

So far this conceptual discussion compares quasi-static states. The complexity of the DW motion under an applied external field can also yield other transitions. Therefore extensive micromagnetic simulations need to be performed to study the influence of the syphon geometry on the DW motion through the cross. The latter is then the key step to finding parameters that minimize failure and optimize reliability.

All the previous discussion was based on an analysis of a Head-to-Head DW with $\alpha$ being between 90$^\circ$ to 0$^\circ$ and 0$^\circ$ to 270$^\circ$. By symmetry, this is equally applicable to a Tail-to-Tail DW with $\alpha$ being between 90$^\circ$ to 180$^\circ$ and 180$^\circ$ to 270$^\circ$.

\section{Micromagnetic simulations}

All the simulations are performed with the free software Mumax3 \cite{Van11} on GPUs. The materials parameters used for Ni$_{81}$Fe$_{19}$ are: Saturation magnetization M$_s$ = 860 kA/m, the exchange constant A = 1.3 $\times$ 10$^{-11} $ J/m, no magnetocrystalline anisotropy and a Gilbert damping of 0.01. The thickness of the material is 30 nm, the width of the stripe is 300 nm, and the cell size was kept below 5$\times$5$\times$15 nm$^3$. The different relevant parts of the structure (cross and syphon) were separately investigated. The separate analysis is possible since the pinning potential in the syphon due to the corners is sufficiently strong to mask any attraction from the center of the cross. The two elements are then effectively decoupled and can be investigated individually in detail. While we have used a transverse DW as a simple representation of a DW in the previous section, the cross-section, and material used in the experiments and the micromagnetic simulations will favor a vortex DW. All simulations are thus started with a vortex DW that corresponds to the experimentally expected wall spin structure.

\subsection{Crossing of two magnetic stripes}

So far, the poorly controlled nature of DW propagation through the cross without a syphon is an effect that limits the use of n-CL structures for sensors.  Despite some studies on the propagation of the DW in split paths \cite{Lew09, Pus13, Bur14, Set16}, DW dynamics propagating through the center of the cross is not yet fully understood, particularly its behavior about the reversal of the vertical arm of the cross.

\begin{figure}[H]
\centering
\includegraphics[scale=0.4]{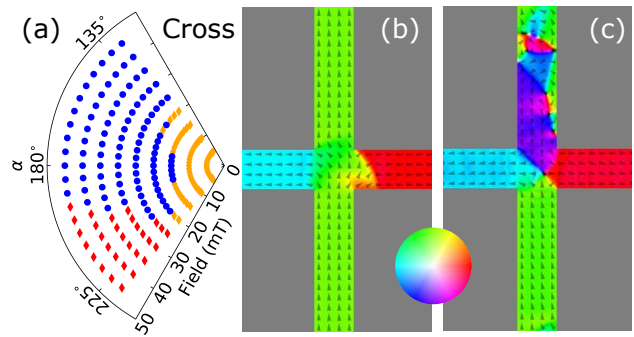}
\caption{ (a) Polar plot of the angular dependence of a Tail-to-Tail DW propagation with field orientation and magnitude represented as radius and angle respectively. The blue circles indicate successful propagation through the cross; the red diamonds represent a reversal of the vertical arm and the orange the pinning at the cross. (b) Spin structure of a DW pinned at a cross. (c) Snapshot of magnetization dynamics showing the splitting of the DW generating the reversal of the vertical arm in addition to the horizontal arm and thus representing a failure event. The in-plane magnetization direction is indicated by the color code visible in the color wheel.}
\label{fig5}
\end{figure}

A systematic study of the DW propagation has been performed changing both the applied field strength and orientation. An initial configuration was obtained by placing a Tail-to-Tail vortex DW on the left arm and letting the system relax under no external applied field. A field with a chosen direction and magnitude was then applied, and the system was left to relax. Different testing conditions were then probed to determine if the desired behavior was obtained, namely the motion from left to right through the cross without the reversal of a vertical arm. 

The orange circles represent the pinning of the DW in the center (Fig. \ref{fig5}-b), the red diamonds the reversal of the vertical arm (Fig. \ref{fig5}-c), and the blue diamonds the successful propagation through the cross (Fig. \ref{fig2} and \ref{fig3}-b). We notice that since the domain wall is Tail-to-Tail, the angular range for $\alpha$ is between 90$^\circ$ and 270$^\circ$ (centered around 180$^\circ$) to be able to propagate the wall from left to right. The simulations are performed for the angle range between 130$^\circ$ and 230$^\circ$ which is sufficient to observe all the interesting switching behavior. The field amplitude is also kept below 45 mT, which is the nucleation field value of a wire in our system extracted from the Stoner – Wohlfarth model \cite{SW48}. Indeed, if the field reaches higher values, DWs can nucleate in the closed system, destroying the two studied spin structures. The applied field can be rotated clockwise or counter-clockwise in the device. 
However, from symmetry arguments, it is sufficient to scan one direction, and so we scan the plot starting from 240$^\circ$ towards 110$^\circ$ for different magnitudes of the field. 
At the first data points (230$^\circ$), only failure events are encountered, i.e. pinning below 20 mT and vertical arm reversal above. The first propagation is reached for 20 mT at 220$^\circ$. 
Regarding fields strengths, the first propagation occurs at 15 mT for a range of 10$^\circ$ centered around 180$^\circ$ denoting that the lowest depinning field is in between 10 and 15 mT (not observable due to our field step size). We observe an increase of the angular range of propagation as the field strength increases. 
This tendency continues until 25 mT, where the first reversal of the vertical arm occurs. 
The depinning dependence appears not symmetrical around the x-axis due to the smaller applied torque to the DW at lower angles while it is pinned in the center of the cross. At the entrance of the cross, the abrupt change in magneto-statics constitutes a potential well for the DW. The DW in the wire is confined to a pseudo-one dimensional structure and keeps a well-defined vortex wall internal structure with two edge defects. 
At the cross,  however, due to the lack of vertical confinement, one of the two edge defects annihilates and renucleates on the opposite side of the horizontal arm while the other $\frac{1}{2}$-defect is spread across the vertical arm effectively reducing the stray field (seen Fig. \ref{fig5}-b). The fact that the half edge defect present in the stripe is always the one with the lower stray field thus supports our interpretation. 

Above 25 mT, the angular range decreases due to the large vertical component of the applied field, which promotes the propagation along the vertical arm. At 45 mT, an angular range of 10$^\circ$ for the propagation is still present in the bottom quadrant. No reversal of the vertical arm is observed in the top quadrant because the vertical component of the applied field is parallel to the magnetization. Concerning the vertical arm reversal, the variety of spin configurations \cite{ NLS74, Zei11} that the DW can acquire at high fields while interacting with the center of the cross makes the dynamics complex.

\begin{figure}[H]
\centering
\includegraphics[scale=0.37]{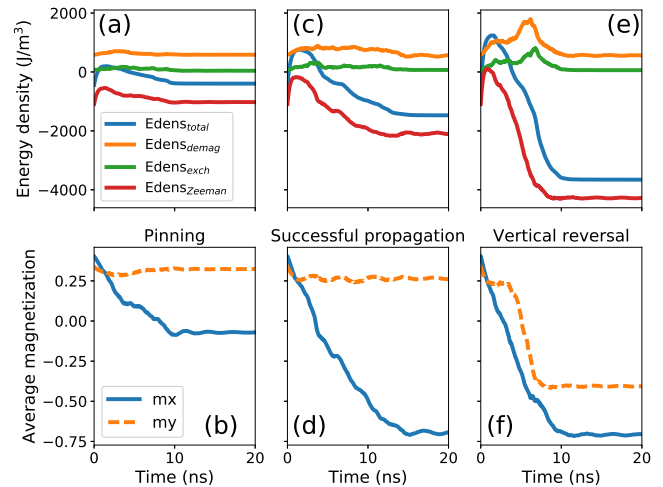}
\caption{Time evolution of (a) the energy density terms and (b) x and y components of the magnetization in the case of the DW being pinned at the cross (Field: $\alpha$ = 210$^\circ$, strength = 10 mT). Time evolution of (c) the energy density terms and (d) x and y components of the magnetization in the case of a DW propagating through the cross (Field: $\alpha$ = 210$^\circ$, strength = 20 mT). Time evolution of (d) the energy density terms and (e) x and y components of the magnetization in the case of a DW inducing vertical arm reversal (Field: $\alpha$ = 210$^\circ$, strength = 30 mT).}
\label{fig6}
\end{figure}

The average magnetization and the energy density terms evolution across time is shown in Fig. \ref{fig6} for the three characteristic events described previously.
As expected, the average x-component is larger for the pinning case than for the two others due to the non-complete reversal of the horizontal arm. Similarly, a strong change of the average y-component is observable in the vertical reversal case. Considering the energy density terms, the total energy density is mainly driven by the Zeeman energy density of the system. In the three cases, an increase in the Zeeman energy density, first, occurs due to the ramp up of the applied field that follows a functional dependence of the form $ B = (B_{ext}(1-exp(\frac{-t}{5e-10}))cos(\alpha)$, $B_{ext}(1-exp(\frac{-t}{5e-10}))sin(\alpha)$,0). The following decrease is due to the DW movement switching the magnetization to a lower energy state. The magnitude of the Zeeman energy density at equilibrium is then directly proportional to the amount of magnetic material switched in the direction of the applied field. For the pinned case, half the horizontal wire switched, for the successful propagation, the whole horizontal wire switched, and for the vertical reversal the entire system switched. A movie for the 3 processes is provided in the supplemental material \cite{supp}.

Due to the rotating field used in the actual sensor device concept, if the vertical arm has a magnetization anti-parallel to the applied field, the reversal of the vertical arm is likely to occur resulting in a failure event. For example, if a Tail-to-Tail DW is found to the left of the cross when a field of 40 mT is applied along 200$^\circ$, according to Fig. $\ref{fig5}$-a, its propagation will lead to a reversal of the vertical arm. To overcome this issue, we next introduce additionally the syphon element, which will constrain the propagation through the cross at applied field angles close to 180$^\circ$.

\subsection{Syphon element}

The syphon structure comprises of two horizontal wire segments connected by a tilted wire at an angle $\theta$ from the horizontal direction (Fig. \ref{fig4}). The goal is to limit the movement of the DW to field directions that lead to successful horizontal DW propagation without vertical DW propagation as represented by the blue area in Fig. $\ref{fig5}$-a. The syphon arms are placed on each side of the cross to allow for the use of the device with a clockwise and counter-clockwise rotating field. 

\begin{figure}[H]
\centering
\includegraphics[scale=0.35]{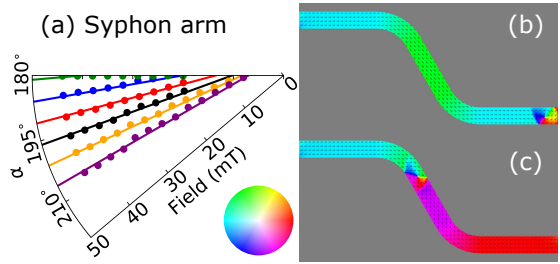}
\caption{ (a) Polar plot depicting the limit between pinning and propagation at the syphon arms. The circles are simulation results for the syphon angle $\theta$ = 85$^\circ$, 80$^\circ$, 75$^\circ$, 70$^\circ$, 65$^\circ$ and 60$^\circ$, respectively in green, blue, red, black, orange and purple. The lines are the fitting curves of the simulated data based on equation 1. Taking the results for $\theta$ = 60$^\circ$ (purple), the DW pinning occurs for field angles larger than 205$^\circ$, the propagation is allowed for field angles between 180$^\circ$ and 205$^\circ$ and field strengths larger than 7 mT. (b) Successful propagation of the DW through the syphon. (c) Pinning at the center of the syphon. The in-plane magnetization direction is indicated by the color code visible in the color wheel.}
\label{fig7}
\end{figure}

In simulations, a DW is initialized at the top-left horizontal arm of the syphon and propagated towards the right. The two final configurations are the desired propagation shown in Fig. \ref{fig7}-b or the pinning of the DW in the center of the arm (Fig. \ref{fig7}-c). Different syphon angles ($\theta$ in Fig. \ref{fig4}-b) were simulated. The results can be seen on the polar plot in Fig. \ref{fig7}-a. In Fig. \ref{fig7} -a, we present only the limit between pinning and propagation. The circles are the simulation results while the lines are the fitting of the results for an angle of the syphon arm  (Equation 1) with $\theta$ varying from 85$^\circ$ to 60$^\circ$ in steps of 5$^\circ$ (left syphon arm). 

The results for a syphon with $\theta$ = 60$^\circ$, represented in purple,  show that the domain wall will propagate through the syphon for field values between 7 mT and 45 mT and angle values between 180$^\circ$ and 205$^\circ$ (purple line). However, the domain wall will be pinned for applied fields at any angle higher than the limit marked by the solid line. The other syphon angles exhibit similar behavior but with different limits.

We study the left arm of the syphon since all results can be applied to the right syphon by reflection symmetry along the vertical axis. Similarly, due to the symmetry of the system, if the angle of the applied field is smaller than $\alpha$ = 180$^\circ$, the DW will propagate through the first syphon arm and the cross but not across the second syphon arm. 
Thus we are then only interested in the angles between $\alpha$ = 180$^\circ$ and 220$^\circ$. From the results, we find that the angular range increases with the magnitude of the applied field and the reduction of the syphon angle ($\theta$).

The propagation of the DW through the syphon element is described by a model of a DW in a straight wire under the application of a field with axial and transverse components \cite{Bry08, Gla08}.
In a 1D model \cite{Thi02, Nak05}, the torque on the DW from the applied  field can be approximated as $\mu_0 H_{ext} M_s \sin(\frac{3\pi}{2}-\theta-\alpha)$ and it is maximum when the applied field is parallel and opposite to the magnetization in the stripe. From the latter, we extract the depinning field in the arm:

\begin{equation}\label{eq:1}
H_{ext} = \frac{H_{p}}{|sin(\frac{3\pi}{2}-\theta-\alpha)|} 
\end{equation}

With $H_{p} $ the pinning field in the structure due to the corners geometry. For $H_{p} $ = 3.5 mT, we notice the excellent agreement with the model. The syphon arm can thus prevent the DW from propagating for field directions outside the desired angular range. In a real device, the edge roughness might affect the propagation field in the syphon element. The result is likely to be a larger required longitudinal component of the field along the syphon arm to allow for the depinning of the DW. Although, we have not investigated the effect of edge roughness in our opinion the behavior at the cross will not be affected, and that of the propagation fields through the syphon might increase a little, but the angular dependence will remain.

\section{Full device geometry}

\begin{figure}[H]
\centering
\includegraphics[scale=0.4]{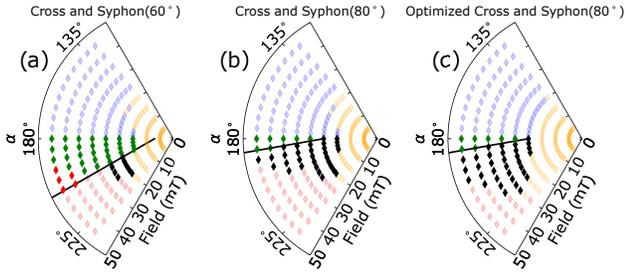} 
\caption{ (a) Polar plot of the combination of a cross with a syphon arm of angle 60$^\circ$ (device n$^\circ$1). The yellow color represents pinning of the domain wall due to a too low field strength. The green color represents the propagation through the full structure; the red circles are the reversal of the vertical arm, and the black region is where the device could potentially still be working if propagation through the syphon was facilitated. We refer to it as a buffer zone. (b) A similar plot for a device with syphon angle 80$^\circ$ (device n$^\circ$2). (c) A similar plot for an optimized device with a cross center of 210 nm and syphon angle 80$^\circ$ (device n$^\circ$3). The black line is the fitting limit expressed previously for the chosen syphon arm. }
\label{fig8} 
\end{figure}

Finally, we now combine both syphon and cross elements for the creation of a reliable device geometry with the desired domain wall propagation properties. The creation of the angular dependence of a complete device is simply accomplished by merging the angular dependences of the two elements. This is justified since the interaction between the elements is small due to the physical distance separating them. The results are shown in Fig. \ref{fig8} where the following scheme has been chosen to represent the results: if the DW propagates across both elements as desired, field direction and strength are represented by a green diamond. If a vertical reversal can still occur then, the diamonds are red. The black diamonds represent the configurations were the DW would be pinned in the syphon arm but if released in the cross would still successfully propagate. We call this the buffer zone. All the irrelevant points of the cross-related effects are now colored with less intensity (faded). A device made of a certain cross and syphon is expected to function properly if no red diamonds (vertical reversal) are seen and if the buffer zone is large enough to account for stochastic events due to thermal activation and irregularities, which would broaden the micro-magnetically computed working range boundaries.

The merging of a cross and a syphon arm with an angle $\theta$ = 60$^\circ$ yields the polar plot in Fig. \ref{fig8}-a (device n$^\circ$1). We note that the field is rotating and that all non-faded points need to be examined to assess whether a device is working or not. 
We notice that in the allowed range of device n$^\circ$1, the application of a field larger than 25 mT would lead to a reversal of the vertical arm since there is no buffer region above it. 
 If we are to define an angle operating window (AOW) between the largest working angle 200$^\circ$ (at 25 mT) and 180$^\circ$, device n$^\circ$1 has an AOW of 20$^\circ$. It is useful to define a field operating window (FOW) between the depinning field limit and the central arm reversal limit. For device n$^\circ$1$  $, the FOW is 10 mT  (15 mT to 25 mT). Finally, the robustness of the structure to processing irregularities and stochasticity can be identified by the smallest black angular range. For device n$^\circ$1, the robustness is of the order of 15$^\circ$ at 25 mT. We believe that for a device, the best parameters are met for a large FOW, a small AOW, and a wide Buffer zone or robustness to irregularities due to device fabrication. A summary of the values for those parameters are found in Table \ref{fig9}.
  
The second plot (Fig. \ref{fig8}-b) represents the characteristics of device n$^\circ$2. Device n$^\circ$2 is created by merging the cross to a syphon arm of 80$^\circ$. This device has a passing field up to the theoretical nucleation field of the structure (no vertical arm reversal). The FOW of device n$^\circ$2 is 25 mT (20 mT to 45 mT) which represents a substantial improvement to device n$^\circ$1. The AOW is also reduced to 10$^\circ$. However, the Buffer zone is only of 5$^\circ$ at 45 mT thus vulnerable to defects. The latter is due to the strong enlargement of the vertical-arm-reversal area as the magnitude of the field increases.

\begingroup
\squeezetable
\begin{table}[H]
\centering\
\begin{tabular}{c|c|c|c|}
\cline{2-4}
\textbf{}                                                        & \textbf{n$^\circ$1} & \textbf{n$^\circ$2} & \textbf{n$^\circ$3} \\ \hline
\multicolumn{1}{|c|}{\textbf{Angle Operating Window ($^\circ$)}} & \textbf{20}                & \textbf{10}                & \textbf{10}                \\ \hline
\multicolumn{1}{|c|}{\textbf{Field Operating Window (mT)}}       & \textbf{10}                & \textbf{25}                & \textbf{25}                \\ \hline
\multicolumn{1}{|c|}{\textbf{Minimum Buffer Zone ($^\circ$)}}    & \textbf{15}                & \textbf{5}                 & \textbf{15}                \\ \hline
\end{tabular}
\caption{Summary of the three defining parameters of the n-CL device for the three presented architectures.}
\label{fig9}
\end{table}

\endgroup

We now optimize the cross with the best syphon angle to enlarge the buffer region. 
The dimensions in the center of the cross have been reduced from 300 nm to 210 nm (device n$^\circ$3). This reduction will increase the nucleation field and potentially also the propagation field. The results of the reduction are seen on the faded points in Fig. \ref{fig8}-c.  The depinning occurs at 5 mT higher values, but the angular range at 45 mT is much larger. Due to the selection of the syphon angle the increase in depinning field has no impact on the performances of the device. The robustness of device n$^\circ$3 is then drastically improved to 15$^\circ$ at 45 mT compare to n$^\circ$2 with no loss of FOW thus providing the necessary reliability to the structure. Depending on the desired parameter range for the FOW, tailored combinations of a syphon and a cross structure allow for the creation of optimum conditions for maximum reliability operation.

\section{Conclusion}

To summarize, we present a new nonvolatile sensor concept based on magnetic DWs that will allow for counting to millions of turns of an applied external rotating magnetic field. 
The concept requires the use of the cross-shaped intersections of nanowires forming intertwined loops. 
We explore the DW propagation in this cross-shaped geometry and obtain different states of the cross while interacting with a domain wall. We find that with a simple cross, some configurations (applied field and magnetization in the vertical arm being anti-parallel) are generating the failure event for the sensor operation. 

We develop a syphon structure comprised of a tilted wire to overcome the problems leading to failure events. 
We simulate elements under fields of different strengths and angles to identify their operating conditions. The cross yields a complex DW behavior characterized by three physical mechanisms: the propagation through the cross, the pinning at the cross and the reversal of a vertical arm. The syphon arm for different syphon angles is modelled via a hyperbolic dependence on the angle of the applied field. The syphon structure successfully allows for the pinning of the DW in angular ranges where the applied field configuration in correspondence to magnetization state in the cross would yield a failure if the domain wall was to propagate into the cross region at these field directions and field strengths.

Finally, the combination of the two elements allows for the identification an angular operating window, a field operating window and a robustness factor; the three of them allow for an easy identification of a reliable structure, which is essential to design the geometries of reliable devices of this new generation of multi-turn sensor.

\section*{Acknowledgements}
The authors would like to acknowledge K. Litzius for helping with the simulations and the WALL project for financial support. The work and results reported in this publication were obtained with research funding from the European Community under the Seventh Framework Programme - The people Programme, Multi-ITN “WALL” Contract Number Grant agreement no.: 608031, a European Research Council Proof-of-Concept grant (MultiRev ERC-2014-PoC (665672)), the German research foundation (SFB TRR173 Spin+X), and LL-D acknowledges the support by Project SA090U16 from Junta de Castilla y Leon.

%\bibliographystyle{apsrev4-1}
%\bibliography{syphonpaper}

\end{document}